\documentclass[cits]{PoS}

\title{Nova Framework: A New Tool For Modeling of Nova Outbursts and Nucleosynthesis}

\author{\speaker{Pavel A. Denissenkov}$^a$, Falk Herwig$^a$, Marco Pignatari$^b$, James W. Truran$^c$\\
\llap{$^a$} Department of Physics \& Astronomy\\ 
            University of Victoria, P.O.~Box 3055\\
            Victoria, B.C., V8W~3P6, Canada\\
\llap{$^b$} Department of Physics\\ 
            University of Basel, Klingelbergstrasse 82\\
            CH-4056 Basel, Switzerland\\
\llap{$^c$} Department of Astronomy and Astrophysics\\
            University of Chicago\\
            Chicago, IL, USA\\
            E-mail: \email{pavelden@uvic.ca}, \email{fherwig@uvic.ca}, \email{mpignatari@gmail.com}, 
            \email{truran@nova.uchicago.edu}}

\abstract{Classical novae are the results of surface thermonuclear explosions of H-rich material
          accreted by white dwarfs (WDs) from their low-mass main-sequence or red-giant binary companions.
          Chemical composition analysis of their ejecta shows that nova outbursts
          occur on both carbon-oxygen (CO) and oxygen-neon (ONe) WDs, and
          that there is cross-boundary mixing between the accreted envelope and underlying WD.
          We have combined the stellar evolution code MESA and post-processing nucleosynthesis tools of NuGrid
          into a framework that allows to produce up-to-date models of nova outbursts and 
          compute detailed nucleosynthesis in novae occurring on CO and ONe WDs.
          The convective boundary mixing (CBM) in our 1D numerical simulations is implemented using
          a diffusion coefficient that is exponentially decreasing with a distance below the bottom of
          the convective envelope. This MESA CBM prescription is based on the findings in 3D hydrodynamic simulations
          that the velocity field, and along with it the mixing expressed in terms of a diffusion coefficient,
          decays exponentially in the stable layer adjacent to a convective boundary.
          The framework can also use the commonly adopted
          1D nova model in which the CBM is mimicked by assuming that the accreted envelope has been
          pre-enriched with WD's material.
          In this preliminary report, we present the most interesting new results related
          to CO and ONe nova outbursts that have been obtained with the Nova Framework.}

\FullConference{XII International Symposium on Nuclei in the Cosmos 2012\\
                Cairns, Australia\\
                5th - 10th August 2012}

\ShortTitle{Nova Framework}

\begin{document}

\section{Introduction}

Classical novae are the results of thermonuclear explosions of hydrogen on
the surfaces of white dwarfs (WDs) that accrete H-rich material from their low-mass MS or red-giant binary companions.
Depending  on the initial parameters of the binary system, the primary component can end up its evolution
as a CO or a more massive ONe WD.
The accreted material is heated by gravitational compression and, when the pressure at its base reaches
a critical value, H is ignited under partially degenerate conditions, which leads to a thermonuclear runaway (TNR).
The rapidly increasing temperature and energy generation rate trigger convection in the envelope, where
the hot CNO cycle transforms the stable CNO isotopes into their $\beta^+$-unstable counterparts, mainly
$^{13}$N, $^{14}$O, $^{15}$O, and $^{17}$F. From this moment on, the energy generation rate is limited by
the half-life times of the unstable isotopes, of order $10^2$\,--\,$10^3$ seconds.
The degenerate conditions are lifted soon after the temperature starts to rise during the early stages of
the TNR. Mass loss is driven by the TNR and results from the conversion of energy released by nuclear reactions into 
kinetic energy, leading to $v_{\rm env} > v_{\rm esc}$. This will be supplement by additional mass loss from a wind (mainly).
The observed enrichment of the ejecta of novae in heavy elements is believed to be a signature of
mixing between the accreted envelope and WD.
Recent two- and three-dimensional nuclear-hydrodynamic simulations of a nova outburst have shown that
this mixing is most likely to be driven by the hydrodynamic instabilities
and shear-flow turbulence induced by steep horizontal velocity gradients at the bottom of the convection zone that
develops during the explosion \cite{cea11a,cea11b}. Therefore, we call it the convective boundary mixing (CBM).
The large mass fractions, on average 30\% \cite{gea98}, of CNO and Ne nuclei in the ejecta
confirm that nova outbursts and CBM occur on both CO and ONe WDs.

Recently, we have demonstrated that the state-of-the-art stellar evolution code
MESA\footnote{MESA is an open source code; it is described in detail in \cite{pea11} and can be downloaded
from {\bf http://mesa.sourceforge.net}.}
can successfully be used for 1D numerical simulations of nova outbursts occurring on CO WDs \cite{dea12}.
In a forthcoming paper, we will show that MESA can model the evolution of ONe novae
as well. In addition, the multi-zone parallel code MPPNP, which is a part of the NuGrid
research collaboration ({\bf www.nugridstars.org}), is used in the second paper 
for post-processing computations of nucleosynthesis in CO and ONe novae. Our
MESA nova models serve as a background that provides MPPNP with necessary stellar structure and mixing parameter profiles.
In both MESA and NuGrid codes, convection is not approximated by instantaneous mixing but instead treated as 
a diffusion process with diffusion coefficients proportional to local values of convective velocity and mixing length.
As a result of our cooperation with JINA and TRIUMF,
we have combined MESA and NuGrid tools into a framework that allows to produce up-to-date models of nova outbursts and 
nucleosyntheis. The interaction between different parts of the network is controlled by user-friendly shell scripts. 
It can be used for research and educational purposes.
In this preliminary report, we present the most interesting new results related
to CO and ONe nova outbursts that have been obtained with the Nova Framework.

\section{Results Related to CO Novae}

In the first paper \cite{dea12}, we have used the MESA stellar evolution code to construct
multicycle nova evolution sequences with CO WD cores. We have explored a range of WD
masses ($0.65\,M_\odot$, $0.85\,M_\odot$, $1.0\,M_\odot$, $1.15\,M_\odot$, and
$1.2\,M_\odot$) and accretion rates (from $10^{-11}M_\odot/\mbox{yr}$ to
$10^{-9}M_\odot/\mbox{yr}$) as well as the effect of different cooling times (different initial WD temperatures
and luminosities) before the onset of accretion. In addition, we have studied the dependence on the
elemental abundance distribution of accreted material and convective boundary mixing (CBM)
at the core-envelope interface. In the first case, which is commonly adopted in 1D nova
simulations, the CBM was mimicked by assuming that the accreted envelope had been
pre-enriched (pre-mixed) with 30\% of WD's material. In the second case, the CBM was implemented using a
diffusion coefficient that was exponentially decreasing with a distance below the bottom of
the convective envelope. This 1D approximation for CBM is supported by 3D hydrodynamic
simulations of He-shell flash convection in AGB stars \cite{hea11}. Our nova models with
such CBM display an enrichment of the accreted envelope with C and O from the underlying white
dwarf that is commensurate with observations.  We have compared our results with the previous
work and investigated a new scenario for novae with the $^3$He-triggered convection. We have
found that $^3$He can be produced \emph{in situ} in H-rich envelopes accreted with slow rates 
($\dot{M} < 10^{-10}M_\odot/\mbox{yr}$) by cold ($T_{\rm WD} < 10^7$ K) CO WDs, and
that convection is triggered by $^3$He burning before the nova outburst in this case.

\section{Results Related to ONe Novae}

\begin{figure}
\centering
\includegraphics[bb=60 225 480 695]{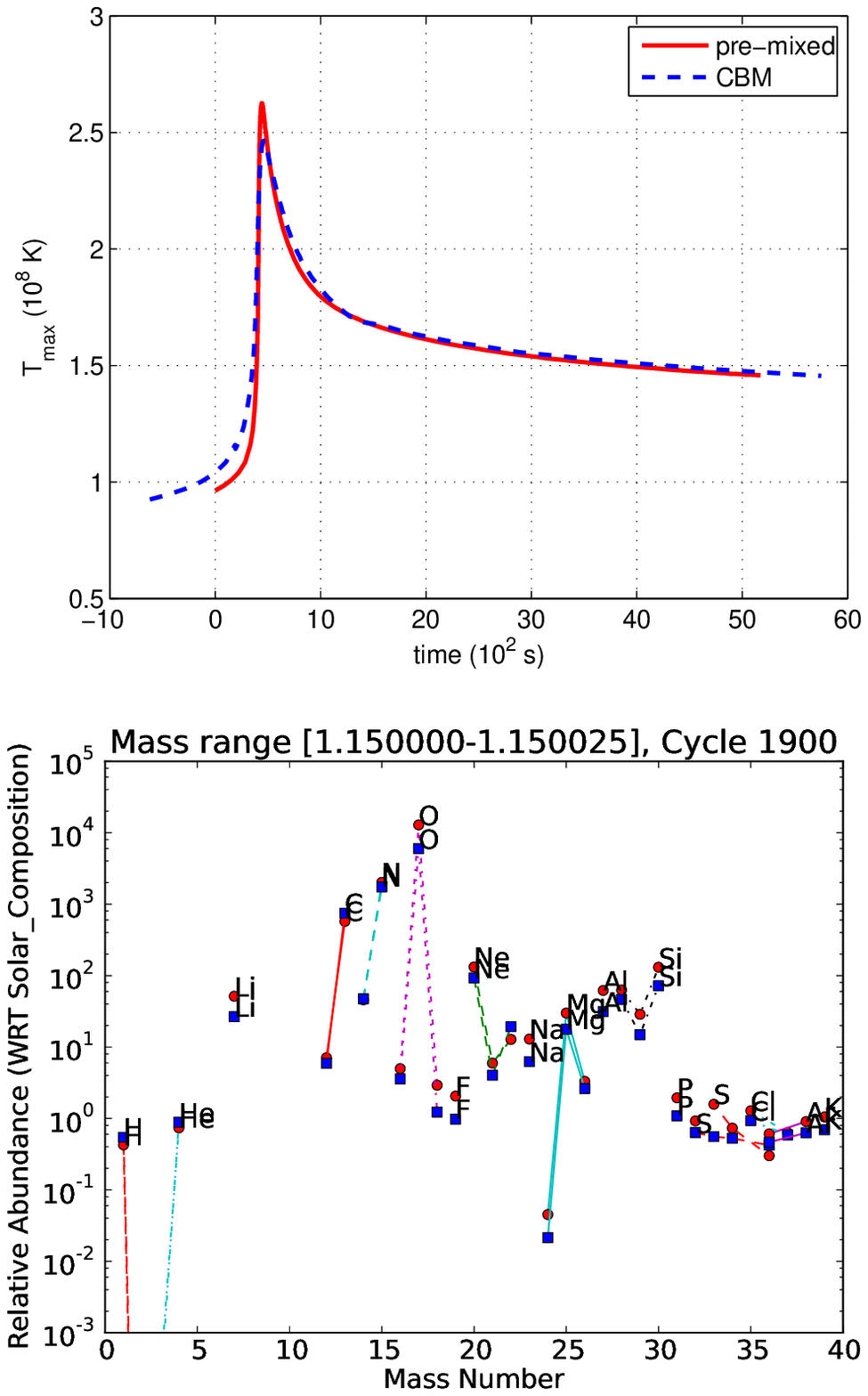}
\caption{A comparison of $T_{\rm max}$ evolution profiles (the upper panel) and final abundances (the lower panel) from two of our
         $1.15\,M_\odot$ ONe nova simulations with
         $T_{\rm WD} = 12$ MK and $\dot{M} = 2\times 10^{-10}\,M_\odot/\mbox{yr}$.
         The solid red curve and red circles are the results obtained with the 50\% pre-mixed
         initial abundances in the accreted envelope (a mixture of equal amounts of WD's and solar-composition
         materials), while the dashed blue curve and blue squares represent the case
         with the solar-composition accreted material and convective boundary mixing (CBM) modeled using the same prescription
         (the exponentially decaying diffusion coefficient) that we used in the $1.2\,M_\odot$ CO nova model in \protect\cite{dea12}.}
\label{fig:f1}
\end{figure}

In the forthcoming paper,
we show that for ONe novae the CBM prescription with the exponentially decaying
diffusion coefficient produces maximum temperature evolution
profiles and nucleosynthesis yields in good agreement with those obtained using the pre-mixed          
nova model (Fig.~\ref{fig:f1}) and that this is not true for CO novae.

\begin{figure}
\centering
\includegraphics[bb=75 200 480 695]{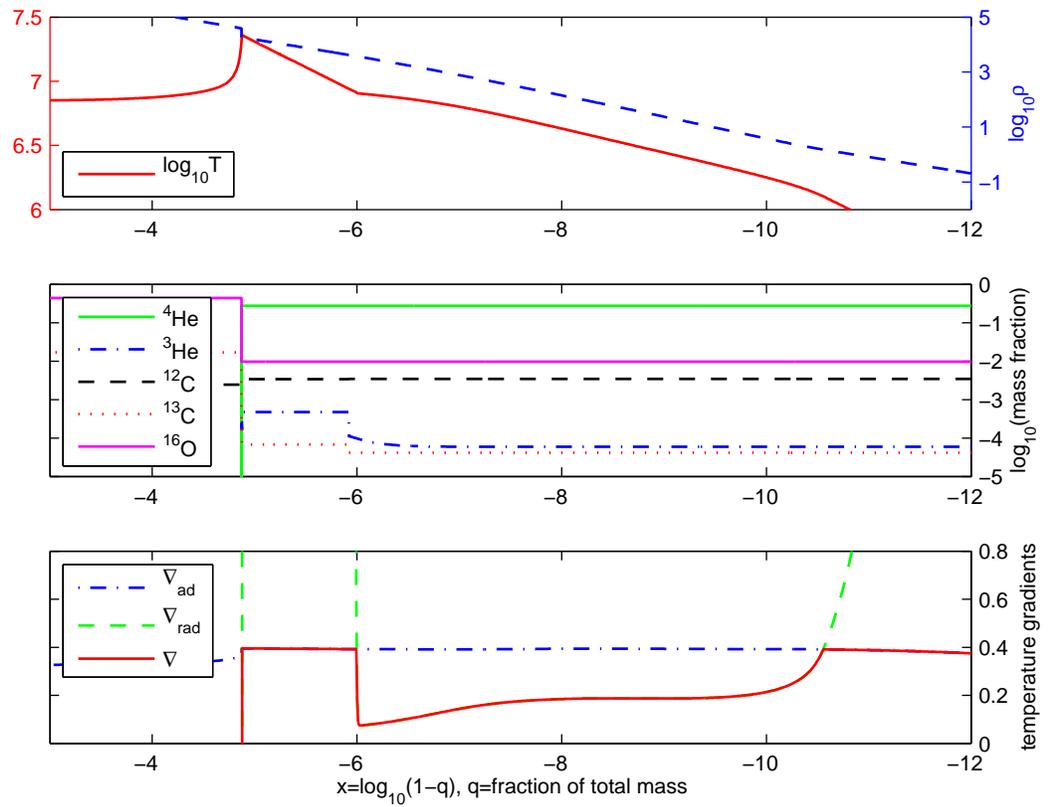}
\caption{A snapshot of profiles of various stellar structure parameters in the envelope of our $1.3\,M_\odot$ ONe nova model with
         $T_{\rm WD} = 7$ MK and $\dot{M} = 10^{-11}\,M_\odot/\mbox{yr}$ at the moment following the $^3$He ignition that
         has triggered convection. Like in the CO nova model with the cold WD and low accretion rate
         discussed in \protect\cite{dea12}, the low $T_{\rm WD}$ and $\dot{M}$ values favour the accumulation of $^3$He in a slope
         adjacent to the core-envelope interface followed by its ignition at a relatively low $T$,
         before the major nova outburst ensues.}
\label{fig:f2}
\end{figure}

We demonstrate that in ONe novae $^3$He can also be produced \emph{in situ}
in H-rich envelopes accreted with slow rates ($\dot{M} < 10^{-10}\,M_\odot/\mbox{yr}$) by cold ($T_{\rm WD} < 10^7$ K)
WDs, and that convection is triggered by $^3$He burning before the nova outburst in this case (Fig.~\ref{fig:f2}).

\begin{figure}
\centering
\includegraphics[bb=75 200 480 695]{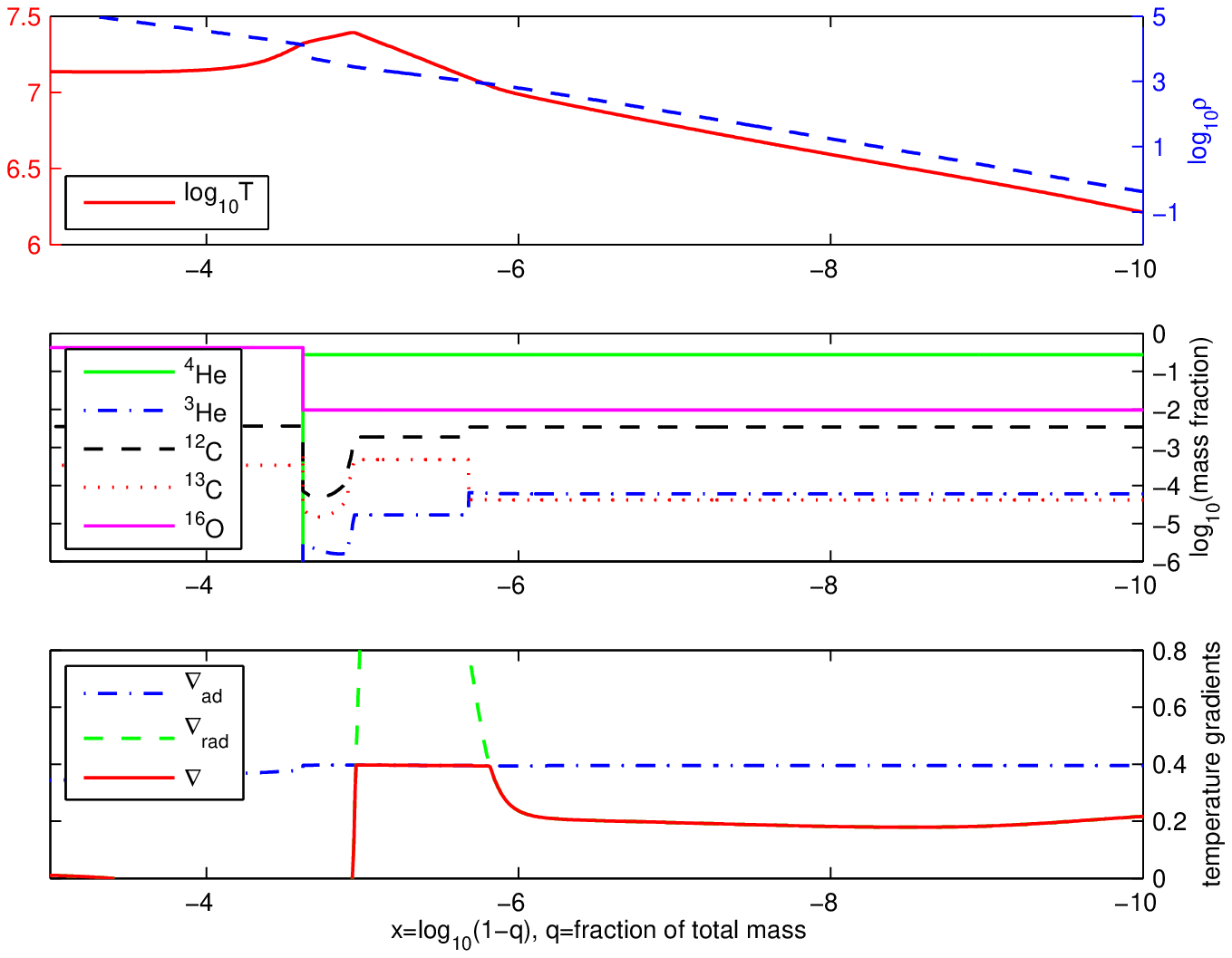}
\caption{A snapshot similar to that shown in Fig.~\protect\ref{fig:f2}, but for our $1.15\,M_\odot$ ONe nova model with
         $T_{\rm WD} = 15$ MK and $\dot{M} = 10^{-10}\,M_\odot/\mbox{yr}$. The interplay between the $^3$He production
         and destruction shifts the peak temperature away from the core-envelope interface, located at the leftmost steps on
         the abundance profiles (the middle panel), to a place where the $^3$He
         burning generates the maximum energy. This leads to the formation of a convective zone separated from
         the interface by a buffer zone.}
\label{fig:f3}
\end{figure}

Additionally, we find that the interplay between the $^3$He
production and destruction in the H-rich envelope accreted with the intermediate rate,
$\dot{M} = 10^{-10}\,M_\odot/\mbox{yr}$, by the $1.15\,M_\odot$ ONe WD with the central temperature
$T_{\rm WD} = 15\times 10^6$ K
leads to the formation of a thick radiative buffer zone that separates the bottom of the convective
envelope from the WD surface (Fig.~\ref{fig:f3}).

\acknowledgments
The speaker appreciates helpful discussions with Lars Bildsten, Barry Davids,
Ami Glasner, Bill Paxton, Chris Ruiz, and Michael Wiescher.

\end{document}